\documentstyle[12pt]{article}
\newcommand{\postscript}[2]{\vspace{5cm}}

\newcommand{\beq}{\begin{equation}}
\newcommand{\eeq}{\end{equation}}
\newcommand{\beqa}{\begin{eqnarray}}
\newcommand{\eeqa}{\end{eqnarray}}
\def\ra{\rightarrow}

\def\etal{{\it et al.}}
\newcommand{\RMP}[3]{{\em Rev. Mod. Phys.} {\bf #1}, #2 (19#3)}
\newcommand{\PR}[3]{{\em Phys. Rev.} {\bf #1}, #2 (19#3)}
\newcommand{\PL}[3]{{\em Phys. Lett.} {\bf #1}, #2 (19#3)}

\newcommand{\PRL}[3]{{\em Phys. Rev. Lett.} {\bf #1}, #2 (19#3)}
\newcommand{\NP}[3]{{\em Nucl. Phys.} {\bf #1}, #2 (19#3)}

\newlength{\dinwidth}
\newlength{\dinmargin}
\begin{document}
{}~~\\
\vspace{1cm}
\begin{center}
\begin{Large}
\begin{bf}
%
%
THEORETICAL STUDY OF THE ELECTROWEAK INTERACTION --
PRESENT AND FUTURE\footnote{The results presented here are based
in part on an analysis in collaboration with Jens Erler, in progress.}\\
\end{bf}
\end{Large}
\vspace{5mm}
\begin{large}
%
%
PAUL LANGACKER \\
\end{large}
%
%
University of Pennsylvania,  Department of Physics \\
Philadelphia, Pennsylvania, USA 19104-6396 \\
\vspace{5mm}
\end{center}
\begin{quotation}
\noindent
%
%
{\bf Abstract:}
There have been several important recent developments in precision
electroweak tests.  These include: the new LEP energy scan during the 1993
run; the first high-precision results on the left-right asymmetry from the
SLD Collaboration at SLAC; the probable discovery of the top quark by the
CDF Collaboration at Fermilab and the determination of its mass.  I will
discuss the implications of these and earlier results for testing the
standard model; for the standard model parameters, including the top quark
mass, the Higgs mass, the weak mixing angle, and the strong coupling
constant, $\alpha_s$; and the search for new physics.  In particular, given
the CDF direct determination of $m_t$ it is now possible to severely
constrain certain types of new physics by separating the contribution from
new physics from the dependence on $m_t$.
\end{quotation}

\section{Introduction}

\begin{itemize}
\item The Two Paths: Unification or Compositeness
\item Recent Data
\item Radiative Corrections
\item Results: $m_t$, $M_H$, $\alpha_s$, $\sin^2 \theta_W$
\item New Physics
\end{itemize}

\section{The Two Paths: Unification or Compositeness}

Most work in particle physics today is directed towards searching for the
new physics beyond the standard model.  Although there are many theoretical
ideas for the nature of such new physics most possibilities fall into one
of two general categories.

The first, which I describe as the Bang scenario, involves the unification
of the interactions.  In such schemes there is generally a grand desert up
to a grand unification (GUT) or Planck scale ($M_P$).  This is the natural
domain of elementary Higgs fields, supersymmetry, GUTs, and superstring
theories.  If nature should choose this route there is a possibility of
probing to $M_P$ and to the very early universe.  There are hints from
coupling constant unification that this may be the correct path.  Some of
the implications are that there should be supersymmetry, which can
ultimately be probed by finding the new superpartners at the LHC.
Secondly, one expects to have a light Higgs boson, which acts much like the
standard model Higgs except that it must be lighter than $110 - 150$~GeV,
which should be detectable at the LHC or possibly at LEP 2.  (The
standard model Higgs could be as heavy as 600 $-$ 1000~GeV.)  Finally, a
very important prediction of at least the simplest cases is that one
expects an {\it absence} of deviations from the standard model predictions
for precision electroweak tests, CP violation, or rare $K$ decays, because
of the decoupling of the heavy superpartners.  Of course, it is hard to
take the observed absence of such deviations as compelling evidence for
supersymmetric unification, but they are nevertheless suggestive.  Some
such schemes also lead to predictions for $m_b$, proton decay, neutrino
masses, and occasionally rare decays.

If the coupling constant unification is not just an accident there are very
few new types of physics other than supersymmetry that could be present
without spoiling it (unless two new effects cancel).  These include
additional heavy $Z'$ bosons, gauge singlets, and a small number of
new sequential, mirror, or exotic fermion families.

The other general possibility is the Whimper scenario, in which nature
consists of onion-like layers of matter at shorter and shorter distance
scales.  This is the domain of composite fermions and scalars and of
dynamical symmetry breaking.  Experimental limits imply that
any new layer of compositeness
would have to be strong binding, and is therefore not analogous to previously
observed levels of compositeness.  If nature should choose
this route, then at most one more layer would be accessible to us at the
LHC and future colliders.  Such schemes generally predict significant rates
for rare decays such as $K \ra \mu e$.  This is a generic feature of almost
all such models, and the fact that they have not been observed is a severe
problem for the general approach and has made it difficult to construct
realistic models.  If one somehow evades the problem of rare decays one
still generally expects to see significant effects in LEP and other
precision observables, including new 4-fermi operators,
decrease of the $Z \ra b\bar{b}$ partial width,
and modifications
to $\rho_0$ and to the parameters $S$, $T$, and $U$.
The fact that these have not been seen constitutes an additional serious
difficulty for most such models.  In the future one would also expect to
see new particles and anomalous interactions among gauge bosons.

\section{Recent Data}

Recent results from $Z$-pole experiments are shown in Table~\ref{tab1}.
These include the preliminary results from the 1993 LEP energy scan,
slightly updated from the values  presented at the Moriond
meeting. These are averages from the ALEPH,
DELPHI, L3, and OPAL detectors, including
a proper treatment of common systematic uncertainties~\cite{a1}.  In
addition, the result from the SLD experiment at SLAC~\cite{a2} on the
left-right asymmetry $A_{LR}$ is shown.  The first row in Table~\ref{tab1}
gives the value of the $Z$ mass, which is now known to remarkable
precision.  Also shown are the lineshape variables $\Gamma_Z$, $R$, and
$\sigma_{\rm had}$; the heavy quark production rates; various
forward-backward asymmetries, $A_{FB}$; quantities derived from the $\tau$
polarization $P_{\tau}$ and its angular distribution; and the effective
weak angle $\bar{s}^2_\ell$ obtained from the jet charge asymmetry. $N_\nu$
is the number of effective active neutrino flavors with masses light enough
to be produced in $Z$ decays.  It is obtained by subtracting the widths for
decays into hadrons and charged leptons from the total width $\Gamma_Z$
from the lineshape.  The asymmetries are expressed in terms of the
quantity
\beq A^o_f = \frac{2 \bar{g}_{Vf} \; \bar{g}_{Af}}{\bar{g}^2_{Vf}
+ \bar{g}^2_{Af}},  \label{eqn1} \eeq
where $\bar{g}_{V,Af}$ are the vector and axial vector couplings to fermion
$f$.

\begin{table} \centering 
\begin{tabular}{|l|c|c|}
\hline \hline
Quantity & Value & Standard Model \\ \hline
$M_Z$ (GeV) & $91.1895 \pm 0.0044$ &  input \\
$\Gamma_Z$ (GeV) & $2.4969 \pm 0.0038$ & $2.496 \pm 0.001 \pm
0.003 \pm [0.003] $ \\
$R = \Gamma({\rm had})/\Gamma(\ell \bar{\ell})$ & $20.789 \pm
0.040$ & $20.782 \pm 0.006 \pm 0.004 \pm [0.03]$ \\
$\sigma_{\rm had} = \frac{12 \pi}{M_Z^2} \; \frac{\Gamma(e
\bar{e}) \Gamma({\rm had})}{\Gamma_Z^2} ({\rm nb})$ & $41.51 \pm 0.12$
& $41.45 \pm 0.01 \pm 0.01 \pm [0.03]$ \\
$R_b = \Gamma(b \bar{b})/ \Gamma({\rm had})$ &$0.2208 \pm 0.0024$
& $0.2155 \pm 0 \pm 0.0004$ \\
$R_c = \Gamma(c\bar{c}) /\Gamma({\rm had})$ & $0.170 \pm 0.014$
 & $0.171 \pm 0 \pm 0$ \\
$A^{0\ell}_{FB} = \frac{3}{4} \left( A_{\ell}^0 \right)^2$ &
$0.0170 \pm 0.0016$ & $0.0152 \pm 0.0005 \pm 0.0007$  \\
$A^0_{\tau} \left(P_\tau \right)$ & $0.150 \pm 0.010$ & $0.142
\pm 0.003 \pm 0.003$ \\
$A^0_e \left( P_\tau\right)$ & $0.120 \pm 0.012$ & $0.142 \pm
0.003 \pm 0.003$ \\
$A^{0b}_{FB} = \frac{3}{4} A^0_e A^0_b$ & $0.0960 \pm 0.0043$ &
$0.0998 \pm 0.002 \pm 0.002$ \\
$A^{0c}_{FB} = \frac{3}{4} A^0_e A^0_c$ & $0.070 \pm 0.011$ &
$0.071 \pm 0.001 \pm 0.002$ \\
$\bar{s}^2_{\ell} \left(A^{Q}_{FB} \right)$ & $0.2320 \pm
0.0016$ & $0.2321 \pm 0.0003 \pm 0.0004$ \\
$A^0_e \left(A^0_{LR} \right)$ \ \ (SLD) & $0.1637 \pm 0.0075
\;\; (92 + 93)$ & $0.142 \pm 0.003 \pm 0.003$ \\
     & $(0.1656 \pm 0.0076 \;\; (93))$ & \ \\
$N_\nu$ & $2.985 \pm 0.023$ & $3$ \\ \hline
\end{tabular}
\caption{$Z$-pole observables from LEP and SLD compared to their standard
model expectations.  The standard model prediction is based on $M_Z$ and
uses the global best fit values for $m_t$ and $\alpha_s$, with $M_H$ in the
range $60 - 1000$~GeV.}
\label{tab1}
\end{table}

{}From the $Z$ mass one can predict the other observables,
including electroweak loop effects.  The predictions also depend on the top
quark and Higgs mass, and $\alpha_s$ is needed for the QCD corrections to
the hadronic widths.  The predictions are shown in the third column of Table~1,
using the value $m_t = 173 \pm 11 $~GeV obtained for $M_H = 300$~GeV in
a global best fit to all data.  The first uncertainty is from $M_Z$ and
$\Delta r$ (related to the running of $\alpha$ up to $M_Z$), while the
second is from $m_t$ and $M_H$, allowing the Higgs mass to vary in the
range $60 - 1000$~GeV.  The last uncertainty is the QCD uncertainty from
the value of $\alpha_s$.  Here the value and uncertainty are given by
$\alpha_s = 0.124 \pm 0.005$, obtained from the global fit to the lineshape.

The data is in excellent agreement with the standard model predictions
except for two observables.  The first is
\beq R_b = \frac{\Gamma(b\bar{b})}{\Gamma(\rm had)} = 0.2208 \pm
0.0024.\eeq
This is some $2.2 \sigma$ higher than the standard model expectation
$0.2155 \pm 0.0004$.  Because of special vertex corrections, the $b
\bar{b}$ width actually decreases with $m_t$, as opposed to the other
widths which all increase.  This can be seen in Figure~\ref{fig1}.

\begin{figure}
\postscript{/home/pgl/fort/nc/graph/gam/xxrb.ps}{0.6}
\caption{ Standard model prediction for $R_b \equiv \Gamma(b \bar{b}) /
\Gamma( \rm had)$ as a function of $m_t$, compared with the LEP
experimental value.  Also shown are the D0 lower bound of 131~GeV and the
CDF range $174 \pm 16$~GeV.}
\label{fig1}
\end{figure}

It is apparent that $R_b$ favors a small value of $m_t$.  By itself $R_b$
is insensitive to $M_H$.  However, when combined with other observables,
for which $m_t$ and $M_H$ are strongly correlated, the effect is to favor a
smaller Higgs mass.  Another possibility, if the effect is more than a
statistical fluctuation, is that it may be due to some sort of new physics.
Many types of new physics will couple preferentially to the third
generation, so this is a serious possibility.  However, the experimenters
have not completed an analysis of the effects of $Z$ decaying into light
quarks, one of which radiates a gluon which then turns into a $b \bar{b}$.
Until this is separated from the data any conclusions must be preliminary.

The other discrepancy is the value of the left-right asymmetry
\beq A^0_{LR} = A^0_e = \frac{2 \bar{g}_{Ve}
\bar{g}_{Ae}}{\bar{g}_{Ve}^2 + \bar{g}_{Ae}^2} = 0.164 \pm 0.008
\eeq
obtained by the SLD collaboration.  This is some $2.5 \sigma$ higher than
the standard model expectation of $0.142 \pm 0.004$.  This result by itself
favors a large value of the top quark mass, around 240~GeV.  This certainly
is not in good agreement with other observables.  One possibility is that
it is pointing to new physics.  Possibilities here would include $S < 0$,
where $S$ is a parameter describing certain types of heavy new physics
(see Section \ref{stusec}).  In
addition, there are possible tree-level physics such as heavy $Z'$ bosons
or mixing with heavy exotic doublet leptons, $E'_R$, which could
significantly affect the asymmetry.  However, new physics probably cannot
explain all of the discrepancy with the other observables, because
some of the LEP observables measure precisely the same combination of
couplings as does $A_{LR}$\footnote{The relation makes use only of the
assumption that the LEP and SLD observables are dominated by the $Z$-pole.
The one (unlikely) loophole is the possibility of an important
contribution from other sources, such as new 4-fermi operators.  These are
mainly significant slightly away from the pole (at the pole they are out of
phase with the $Z$ amplitude and do not interfere).}. In particular, the LEP
collaborations measured the forward-backward asymmetry for $e^+e^- \ra Z \ra
e^+e^-$, which yields $A^{0e}_{FB} = \frac{3}{4} A^{0\; 2}_e = 0.0158 \pm
0.0035$, implying $A_e^0 = 0.145 \pm 0.016$.  Furthermore, the angular
distribution of the $\tau$ polarization yields $A_e^0 = 0.120 \pm 0.012$.
Combining these results,
\beq A^0_e \left|_{\rm LEP} = 0.129 \pm 0.010 \right. . \eeq
Thus, there is a direct experimental conflict between the LEP and SLD
values of $A_e^0$ at the $2.7\sigma$ level.  One can also consider the
comparison of LEP with SLD if one assumes lepton family universality.  In
that case, from the forward-backward asymmetries into $e$, $\mu$, and
$\tau$ ($A^{0e}_{FB}$, $A^{0\mu}_{FB}$, $A^{0\tau}_{FB}$) one can determine
the leptonic asymmetry $A^{0\ell}_{FB} = 0.0170 \pm 0.0016$, which implies
$A^0_e = 0.1505 \pm 0.0071$.  Combining this with $A^0_e (P_\tau)$ and with
$A^0_\tau (P_\tau) = 0.150 \pm 0.010$ (which is roughly the $\tau$
polarization averaged over angles) one obtains finally
\beq A^0_{\ell} \left|_{\rm LEP} = 0.145 \pm 0.005 \right. , \eeq
which is again $2.0 \sigma$ below the SLD result.

We therefore almost certainly have an experimental conflict.  It may well
be due to a large statistical fluctuation.  If the central value of the SLD
is correct it would also call for new physics when compared with all of the
other observables.

A word is in order concerning the experiments. The quantity
$A_{LR}$ measured by SLD
is perhaps the cleanest single observable: it is obtained as a
ratio from which most radiative corrections and systematic uncertainties
cancel.  It does, however, require an absolute knowledge of the beam
polarization.  It also has a strong sensitivity to $A_e^0$, even for
relatively small statistics, and leads to the most precise single
determination of $A_e^0$.  The LEP results, on the other hand, are based on
the averaging of a number of observations from the four groups and from
various observables.  None  are individually as precise as the SLD result.
On the other hand, the LEP experiments have done an outstanding job, and
the fact that there are so many observables makes it hard to imagine that
any systematic problem in one or a few of them could significantly affect
the overall result.

It will take more time and more statistics to see whether this is just a
fluctuation or whether there is a true discrepancy.  In the meantime it
leaves me with the problem of how to use the results in global fits.  I
will take the view that any discrepancies are statistical
fluctuations, in which case the prescription is to simply combine the
data.  An alternative would be to
multiply the error in the weighted average of $A_e^0$ from LEP and SLD
by a scale factor $S \sim 3.1$, where $S$ is the square root of the
$\chi^2/$df,
to represent the discrepancy.  The latter is the approach favored by
the Particle Data Group.  I will not follow it here, but will indicate its
effects.

There are many other precision  observables.  Some recent ones
are shown in Table~\ref{tab2}.  These include the D0 limit \cite{a2a} $m_t
> 131$~GeV and the value $m_t = 174 \pm 16$~GeV suggested by the CDF
candidate events \cite{a2b}.  There are new observations of the $W$ mass
\cite{a3} from both D0, which has presented a preliminary new value $79.86
\pm 0.40$~GeV, and from CDF, which finds $80.38 \pm 0.23$~GeV.  Combining
these and earlier data one obtains the results shown.  Other
observables include $M_W/M_Z$ from UA2\footnote{I could, of course,
multiply $M_W/M_Z$ by the LEP $M_Z$ and include the result in the $M_W$
average.  (In fact, such a procedure was carried out in the D0 analysis.)
I do not do so because, in principle, it would introduce a correlation
between $M_Z$ and $M_W$.  In practice, the effect is negligible because of
the tiny uncertainty in $M_Z$.}~\cite{a4}, recent results on neutrino
electron scattering from CHARM II~\cite{a5}, and new measurements of $s_W^2
\equiv 1 - M^2_W/M^2_Z$ from the CCFR collaboration at Fermilab~\cite{a6}.
This on-shell definition of the weak angle is determined from deep
inelastic neutrino scattering with small sensitivity to the top quark mass.
The result combined with earlier experiments~\cite{a7} is also shown.  All
of these quantities are in excellent agreement with the standard model
predictions.
\begin{table} \centering 
\begin{tabular}{|c|c|c|}  \hline \hline
Quantity & Value & Standard Model \\ \hline
$M_W$ (GeV) & $80.17 \pm 0.18$ & $80.31 \pm 0.02 \pm 0.08$ \\
$M_W/M_Z (UA2)$ & $0.8813 \pm 0.0041$ & $0.8807 \pm 0.0002 \pm
0.0008$ \\
$Q_W (C_S)$ & $-71.04 \pm 1.58 \pm [0.88]$ & $-72.90 \pm 0.07 \pm
0.05$ \\
$g_A^{\nu e}$ (CHARM II) & $-0.503 \pm 0.018$ & $-0.506 \pm 0 \pm
0.001$ \\
$g_V^{\nu e}$ (CHARM II) & $-0.025 \pm 0.019$ & $-0.038 \pm 0.001
\pm 0.001$ \\
$s^2_W \equiv 1 - \frac{M_W^2}{M_Z^2}$ & $\begin{array}{c} 0.2218
\pm 0.0059 \;{\rm [CCFR]} \\ 0.2260 \pm 0.0048 \; {\rm [All]}
\end{array}$ & $0.2243 \pm 0.0003\pm 0.0015$ \\
$M_H$ (GeV) & $\geq 60$ LEP & $< \left\{ \begin{array}{c} 0
(600), \; {\rm theory} \\ 0 (800), \; {\rm indirect} \end{array}
\right.$ \\
$m_t$ & $> 131$ D0 & $173 \pm 11 ^{+17}_{-19}$ [indirect]
\\
 \ & $174 \pm 16$ CDF & \ \\
$\alpha_s (M_Z)$ & $\begin{array}{cc} 0.123 \pm 0.006 & {\rm LEP
\; event \; shapes} \\ 0.116 \pm 0.005 & {\rm event \; shapes
}+{\rm
low \; energy} \end{array}$ & $\begin{array}{c} 0.124 \pm 0.005
\pm 0.002 \\ \left[ Z \; \; {\rm lineshape}\right] \end{array}$
\\ \hline
\end{tabular}
\caption{Recent observables from the $W$ mass and other
non-$Z$-pole observations compared with the standard model
expectations.  Direct values and limits on $M_H$, $m_t$, and
$\alpha_s$ are also shown.}
\label{tab2}
\end{table}

In the global fits to be described, all of the earlier low energy
observables not listed in the table are fully incorporated.

\section{Radiative Corrections}

In the electroweak theory one defines the weak angle by
\beq \sin^2 \theta_W \equiv \frac{g'^2}{g^2 + g'^2}
\longrightarrow \sin^2 \hat{\theta}_W (M_Z) \;\;\;\;
(\overline{MS})   \label{eq1}\eeq
where $g'$ and $g$ are respectively the gauge couplings of the $U_1$ and
$SU_2$ gauge groups.  Although initially defined in terms of the gauge
couplings, after spontaneous symmetry breaking one can relate the weak angle
to the $W$ and
$Z$ masses  by
\beq M_W^2 = \frac{A^2}{\sin^2 \theta_W} \longrightarrow
\frac{A^2}{\sin^2 \hat{\theta}_W ( 1- \Delta \hat{r}_W)}
\label{eq2} \eeq and
\beq M_Z^2 = \frac{M^2_W}{\cos^2 \theta_W} \longrightarrow
\frac{M^2_W}{\hat{\rho} \cos^2 \hat{\theta}_W (1 - \Delta
\hat{r}_W) } \label{eq3} \eeq
where
\beq A^2 \equiv \frac{\pi \alpha}{\sqrt{2} G_F} = \left( 37.2802 \
{\rm GeV} \right)^2 .\eeq
The first form of equations (\ref{eq1})--(\ref{eq3}) are valid at tree
level.  However, the data is sufficiently precise that one must include
full one loop radiative corrections, which means that one must replace the
quantities by the expressions shown in the last part of equations.  There
are a number of possible ways of defining the renormalized weak angle.
Here I am using the quantity $\sin^2 \hat{\theta}_W(M_Z)$, which is
renormalized according to modified minimal subtraction,
$\overline{MS}$~\cite{a8}.  This basically means that one removes the
$\frac{1}{n-4}$ poles and some associated constants from the gauge
couplings.  In equation (\ref{eq2}) the quantity $\Delta \hat{r}_W$ contains
the finite radiative corrections which relate the $W$ and $Z$ masses, muon
decay, and QED.  The dominant contribution is given by the running of the
fine structure constant $\alpha$ from low energies, where it is defined in
QED, up to the $Z$-pole, which is the scale relevant for electroweak
interactions,
\beq \frac{1}{1 - \Delta \hat{r}_W} \simeq \frac{\alpha (M_Z)}{\alpha} \sim
\frac{1/128}{1/137}. \eeq
There is only a weak dependence on the top quark mass in this scheme,
leading to a value \beq \Delta \hat{r}_W \sim 0.07 \label{eqa6} \eeq
dominated by the running of $\alpha$.  There is a theoretical uncertainty
from the contribution of light hadrons to the photon self-energy diagrams.
This leads to a theoretical uncertainty of $\pm 0.0009$.  This turns out to
be the dominant theoretical uncertainty in the precision electroweak tests
and, in particular, in the expressions relating the $Z$ mass to other
observables.  A similar effect leads to a significant theoretical
uncertainty in $g_\mu - 2$, which will dominate the experimental
uncertainties in the new Brookhaven experiment unless associated
measurements are made of the cross-section for $e^+e^- \ra $~hadrons at low
energies.

Because $m_t$ is so much heavier than the bottom quark mass there
is large $SU_2$ breaking generated by loop diagrams involving the
top and bottom quarks, in particular from the $W$ and $Z$
self-energy diagrams.
There is little shift in the $W$ mass, because that effect is already
absorbed into the observed value of the Fermi constant, so $\Delta
\hat{r}_W$ has no large $m_t$ dependence.  However, the $Z$ mass prediction
is shifted down.  In particular, the quantity $\hat{\rho}$ in
equation~(\ref{eq3}) depends quadratically on $m_t$.  It is given by
\cite{a9} $\hat{\rho} \sim 1 + \rho_t$, where
$\rho_t = \frac{3G_F m_t^2}{8 \sqrt{2} \pi^2}  \sim
0.0031 (m_t/100 \ {\rm GeV})^2$.
  (There are additional contributions from
bosonic loops.)  For $m_t$ in the range 100 -- 200~GeV the effect on
$\hat{\rho}$ can be quite significant.
$\rho_t$ propagates to other observables and generates most of the major
$m_t$ dependence.
(The one exception is the vertex
correction to $Z \ra b{\bar{b}}$ decay.)

{}From the precise value $M_Z = 91.1895 \pm 0.0044$~GeV from LEP one has
\beq \sin^2 \hat{\theta}_W (M_Z)= 0.2318 \pm 0.0005. \eeq
The uncertainty is an order of magnitude smaller than one had prior to the
$Z$-pole experiments at LEP.  The uncertainty from the experimental error
in the $Z$ mass is negligible, of order $0.00003$.  The theoretical
uncertainty $0.0003$ coming from $\Delta \hat{r}_W$ is much larger.  The
largest uncertainty, however, is from $m_t$ and $M_H$, $\sim 0.0004$.  Here
I have used the range of $m_t$ from the global best fit, and
60 GeV $< M_H <$ 1000 GeV.  If one knew $m_t$
one would have a more precise value of the weak angle.  The sensitivity is
displayed in Figure~\ref{fig1b}.  Clearly, one cannot determine the weak angle
from $M_Z$ alone because of the $m_t$ dependence.  One must have either
other indirect observables with a different dependence on $m_t$ or a direct
measurement.  Before discussing other possibilities, I will digress
somewhat on the radiative corrections~\cite{a8}.

\begin{figure}
\postscript{/home/pgl/fort/nc/graph/mt/xxmt.ps}{0.8}
\caption{Values of $\sin^2 \hat{\theta}_W (M_Z)$ as a function of
$m_t$ from various observables.}
\label{fig1b}
\end{figure}

The radiative corrections fall into three categories.  First, there
are the reduced QED corrections, which involve the emission of real photons
and the exchange of virtual photons but do not include vacuum polarization
diagrams.  These constitute a gauge invariant set, but depend on the
details of the experimental acceptances and cuts.  They generally are
removed from the data by the experimenters.  The second class has already
been described.  It is the electromagnetic vacuum polarization diagrams,
which lead to the running from $\alpha^{-1} \sim 137$ at low energies to
$\alpha (M_Z)^{-1} \sim 128$ at the $Z$-pole.  As we have seen this leads
to a significant uncertainty $\Delta \hat{r}_W \sim \Delta \alpha
(M_Z)/\alpha \sim 0.0009$, which can lead to a shift of approximately 3~GeV
in the predicted value of $m_t$.

The electroweak corrections are now quite important.  One must include full
1-loop corrections as well as  dominant 2-loop effects.  The electroweak
corrections include and are dominated by the gauge self-energy diagrams for
the $W$, $Z$, and $\gamma Z$ mixing.  In addition, there are box diagrams
and vertex corrections, which are smaller but which have to be included.
Recently there has been some progress on the dominant 2-loop effects.  In
particular, the dominant terms of order $\alpha^2 m_t^4$ are included.  The
net effect is to replace~\cite{a11}
\beq \hat{\rho} \ra 1 + \rho_t \left[ 1 + \rho_t R \left(
\frac{M_H}{m_t} \right) \right], \eeq
where
\beq \rho_t = \frac{3G_F m_t^2}{8 \sqrt{2} \pi^2}, \eeq
and $R$, which comes from 2-loop diagrams, is strongly dependent on
$M_H$, with $R (0) = 19 - 2 \pi^2$.  There are additional
smaller contributions which must be included in the numerical analysis.

There are also significant mixed QCD-electroweak diagrams, such as
those obtained
by the exchange of the gluon across the quarks in a self-energy diagram.
The dominant contribution
involves top quark loops and is of order $\alpha \alpha_s m_t^2$.
This leads to the replacement
\beq \hat{\rho} \ra 1 + \rho_t \left[ 1 - 2 \alpha_s (m_t)
\frac{\pi^2 + 3}{9 \pi} \right] \sim 1 + 0.9 \rho_t, \eeq
which raises the predicted value of $m_t$ by approximately 5\%.  Recently
there have been discussions and estimates of $t \bar{t}$ threshold
corrections, which are $O(\alpha \alpha_s^2 m_t^2)$.  These have been
estimated using both perturbative \cite{a12} methods and by dispersion
relations~\cite{a13}.  One estimate~\cite{a12} is that the effect is mainly
to shift the scale at which $\alpha_s$ should be evaluated for the $t$
quark loop, namely $\alpha_s (m_t) \ra \alpha_s (0.15 m_t)$.  This is in
good numerical agreement with the dispersion relation estimate. The
threshold estimates have not been included here, but would raise
the predicted values of $m_t$ by $+ 3$~GeV.

\subsection{The Great Confusion: $\sin^2\theta_W$}

There are a number of different definitions of the renormalized weak angle
used in the literature, leading to considerable confusion.  Each of the
definitions has its advantages and disadvantages.  At tree-level there are
several equivalent expressions, namely
\beq  \sin^2 \theta_W = \frac{g^{\prime2}}{g^2 + g^{\prime2}} = 1
- \frac{M_W^2}{M_Z^2} = \frac{\pi \alpha}{\sqrt{2} G_F M_W^2}.
\label{eq10a1} \eeq
The first definition is based on the coupling constants; the last two take
meaning only after spontaneous symmetry breaking has occurred, and
therefore mix in parts of the theory in addition to the gauge vertices.
At higher order one must define a renormalized angle.  One can use the
different expressions in equation~(\ref{eq10a1}) as starting points, and
the resulting definitions differ by finite terms of order $\alpha$, which
also depend on $m_t$ and $M_H$.  This has lead to
considerable confusion (and sometimes heat).

Two common definitions are based on the spontaneous symmetry breaking (SSB)
of the theory, namely on the gauge boson masses.  The most famous is the
on-shell definition~\cite{a8}
\beq s_W^2 = 1 - \frac{M_W^2}{M_Z^2} = 0.2242 \pm 0.0012. \label{7.17}
\eeq
This is very simple conceptually.  However, the $W$ mass is not determined
as precisely as $M_Z$, so $s_W^2$ must actually be extracted from other
data and not from the defining relation (\ref{7.17}).  This leads to a strong
dependence on $m_t$, which accounts for almost all of the uncertainty in
$s_W^2$.  (The value for $s_W^2$ and the other definitions is from a global
fit to all data.)

The $Z$-mass definition \cite{a15},
\beq s^2_{M_Z} \left(1 - s_{M_Z}^2 \right) = \frac{\pi \alpha
(M_Z)}{\sqrt{2} G_F M_Z^2} = 0.2312 \pm 0.0003, \eeq
is obtained by simply removing the $m_t$ dependence from the expression for
the $Z$ mass.  This is the most precise -- the uncertainty is mainly from
$\alpha (M_Z)$.  The use of $s^2_{M_Z}$ is essentially equivalent to using
the $Z$ mass as a renormalized parameter, introducing the weak angle as a
useful derived quantity.  This scheme is simple and precise, and by
definition there is no $m_t$ dependence in the relation betwen $M_Z$ and
$s^2_{M_Z}$.  However, the $m_t$ dependence and uncertainties enter as soon as
one tries to predict other quantities in terms of it.

Both of the definitions based on spontaneous symmetry breaking tend to be
awkward in the presence of new physics, which might shift the values of the
gauge boson masses.  There are other definitions based on the gauge
coupling constants.  These are especially useful for applications to grand
unification, and they tend to be less sensitive to the presence of new
physics.  One is the modified minimal subtraction or $(\overline{MS})$
definition \cite{a8}
\beq \hat{s}^2_Z = \frac{\hat{g}'^2 (M_Z)}{\hat{g}'^2 (M_Z) +
\hat{g}^2 (M_Z)} = 0.2317 \pm 0.0004 ,\eeq
defined by removing the poles and associated constants from the gauge
couplings.  As we have seen, the uncertainty is mainly from $\alpha(M_Z)$
and $m_t$.  There are variant definitions of $\hat{s}^2_Z$, depending on
the treatment of $\alpha \ln (m_t/M_Z)$ terms.  One cannot decouple all
such terms because $m_t \gg m_b$ breaks $SU_2$.  The version used here
\cite{a17} decouples them from $\gamma - Z$ mixing, essentially eliminating
any $m_t$ dependence from the $Z$-pole asymmetry formulas.

Finally, the experimental groups at LEP and SLC have made
extensive use of
\begin{eqnarray} \bar{g}_{Af} &=& \sqrt{\rho_f} t_{3f} \nonumber \\
\bar{g}_{Vf} &=& \sqrt{\rho_f} \left[ t_{3f} - 2 \bar{s}^2_f q_f \right].
\label{eq10a5} \end{eqnarray}
These are the effective axial and vector couplings of the $Z$ to fermion
$f$.  In equation (\ref{eq10a5}) $t_{3f} = \pm \frac{1}{2}$ is the weak
isospin of fermion $f$ and $q_f$ is its electric charge.  The electroweak
self-energy and vertex corrections are absorbed into the coefficient
$\rho_f$ and the effective weak angle $\bar{s}^2_f$. The $\bar{g}_{V,Af}$ are
obtained from the data after removing all photonic contributions.  In
principle there are also electroweak box contributions.  However, these are
very small and are typically ignored or removed from the data.

The effective weak angle differs for different fermions. $\bar{s}^2_{f}$ is
related to the $\overline{MS}$ angle, for example, by
\beq \bar{s}^2_f = \kappa_f \hat{s}^2_Z \eeq
where $\kappa_f$ is a form factor.  The best measured is for the charged
leptons, for which
\beq \bar{s}^2_{\ell} \sim \hat{s}_Z^2 + 0.0002 = 0.2319 \pm
0.0004 \eeq
where there is an additional theoretical uncertainty of $\pm 0.0001$ from
the precise definition of the angles and higher order effects.  These
effective angles are very simple for the discussion of the $Z$-pole data,
but they are difficult to relate to other types of observables.  All
of these definitions have advantages and disadvantages, some of which are
listed in Table~\ref{tab3}.

\begin{table} \centering  
 \begin{tabular}{|l|}  \hline \hline
On-shell : $s^2_W = 1 - \frac{M_W^2}{M_Z^2} = 0.2242\, (12)$ \\
\hline
$+$ most familiar \\
$+$ simple conceptually \\
$-$ large $m_t$ dependence from $Z$-pole observables \\
$-$ depends on SSB mechanism
$-$ awkward for new physics \\ \hline \hline
$Z$-mass : $s^2_{M_Z} = 0.2312 (3)$ \\ \hline
$+$ most precise (no $m_t$ dependence) \\
$+$ simple conceptually \\
$-$ $m_t$ reenters when predicting other observables \\
$-$ depends on SSB mechanism
$-$ awkward for new physics \\ \hline\hline
$\overline{MS}$ : $\hat{s}^2_Z = 0.2317\, (4)$ \\ \hline
$+$ based on coupling constants \\
$+$ convenient for GUTs \\
$+$ usually insensitive to new physics \\
$+$ $Z$ asymmetries $\sim$ independent of $m_t$ \\
$-$ theorists definition; not simple conceptually \\
$-$ usually determined by global fit \\
$-$ some sensitivity to $m_t$ \\
$-$ variant forms ($m_t$ cannot be decoupled in all processes
$(\hat{s}^2_{ND}$ larger by $0.0001 - 0.0002$) \\ \hline \hline
effective : $\bar{s}^2_{\ell} = 0.2319 \pm 0.0004$ \\ \hline
$+$ simple \\
$+$ $Z$ asymmetry independent of $m_t$ \\
$+$ $Z$ widths: $m_t$ in $\rho_f$ only \\
$-$ phenomenological; exact definition in computer code \\
$-$ different for each $f$ \\
$-$ hard to relate to non $Z$-pole observables \\ \hline \hline
\end{tabular}
\caption{Advantages and disadvantages of several definitions of
the weak angle.}
\label{tab3}
\end{table}

\subsection{Other $Z$-Pole Observables}

The other $Z$-pole observables can also be computed.  For example, the
partial width for $Z$ to decay into fermions $f \bar{f}$ is given
approximately by
\beq \Gamma (f\bar{f}) \simeq C_F \frac{G_F M_Z^3}{6 \sqrt{2}
\pi} \left[ \left| \bar{g}_{Af} \right|^2 + \left| \bar{g}_{Vf}
\right|^2 \right].\label{eq12a1}  \eeq
For the heavier quarks and leptons kinematic mass corrections must be
applied.  Effective couplings are proportional to $\sqrt{\hat{\rho}}$ so
that each partial width increases quadratically with $m_t$.  This comes
from the replacement
\beq \frac{M_Z g^2}{8 \cos^2 \hat{\theta}_W}  \ra \hat{\rho}
\frac{G_F}{\sqrt{2}} M_Z^3, \eeq
which incorporates many of the low energy corrections.  In
equation~(\ref{eq12a1}) there is an additional coefficient
\beq  C_f = \left\{ \begin{array}{cc} 1 + \frac{3 \alpha}{4 \pi}
q_f^2 \;\;\;\; , & {\rm leptons} \\ 3 \left( 1 + \frac{3
\alpha}{4 \pi} q_f^2 \right) & \left( 1 + \frac{\alpha_s}{\pi} +
1.409 \left( \frac{\alpha_s}{\pi} \right)^2  -12.77 \left(
\frac{\alpha_s}{\pi} \right)^3 \right) \ \ \ {\rm quarks}
 \end{array} \right. ,
\eeq
which includes QED and QCD corrections.  In particular, the
$\alpha_s$ dependence of the hadronic widths leads to a determination of
$\alpha_s = 0.124 \pm 0.005$ just from the lineshape.  For fixed $M_Z$
most of the $m_t$ dependence is in the $\hat{\rho}$ factor.  One major
exception\footnote{There is also an indirect $m_t$ dependence in
$\bar{s}^2_f$ if one regards $M_Z$ as fixed.} is that $\Gamma (b \bar{b})$
decreases with $m_t$ due to special $m_t$-dependent vertex
corrections~\cite{a18}, \cite{a19}.  These are included in the $\rho_b$ and
$\kappa_b$ factors, but to an excellent numerical approximation
$\Gamma (b \bar{b})$ can be written
as \cite{a19},
\beq \Gamma (b\bar{b}) \ra \Gamma^0(b\bar{b}) \left( 1+ \delta^{SM}_{bb}
\right)\sim \Gamma^0 (b\bar{b}) \left[ 1 - 10^{-2}
\left( \frac{m_t^2}{2M_Z^2} - \frac{1}{5} \right) \right],\label{eq26a}
\eeq
where $\Gamma^0 (b\bar{b})$ is the standard model expression without the
corrections.  This special dependence is useful for separating the
$m_t$ and Higgs effects.

In addition  there are various asymmetries observed at LEP and
SLD.  In particular, the forward-backward asymmetry for $e^+e^-
\ra Z \ra f \bar{f}$ is given, after removing photonic effects and
boxes, by
\beq A^{0f}_{FB} \simeq \frac{3}{4} A_e^0 A_f^0 , \eeq
where $A_f^0$ is defined in (\ref{eqn1}).
Other asymmetries include the polarization of produced $\tau$'s.  From the
angular distribution of the $\tau$ polarization one can obtain
$A_{\tau}^0$ and $A_e^0$, with
$A_{\tau}^0$ coming mainly from the average polorization and $A_e^0$ mainly
from its forward-backward asymmetry.  The SLD collaboration has polarized
electrons; from the left-right asymmetry as the polarization is reversed
one can also determine $A_e^0$, namely $A_{LR}^0 = A_e^0$.

All of these asymmetries are independent of $m_t$ when expressed in terms
of the effective angles $\bar{s}_f^2$ and almost independent of $m_t$ when
expressed in terms of the $\overline{MS}$ angle $\hat{s}^2_Z$.  One can
therefore determine $\bar{s}_\ell^2$ or $\hat{s}^2_Z$ from the data without
theoretical uncertainties from $m_t$.  On the other hand, in the on-shell or
$Z$-mass schemes the formulas involve quadratic $m_t$ dependence.

\section{Results: $m_t$, $M_H$, $\alpha_s$, $\sin^2\theta_W$}

There are now sufficiently many observables that one can
precisely determine $\hat{s}^2_Z$, $m_t$, and $\alpha_s (M_Z)$
simultaneously.  For example, $\hat{s}^2_Z$ can be determined
from the asymmetries,  $m_t$ from the $W$ and $Z$ masses, and
$\alpha_s (M_Z)$ from the hadronic $Z$-widths.  In practice all
of these quantities are determined from a simultaneous fit.  The
results of fits to various sets of data are shown in
Table~\ref{tab4}.
\begin{table} \centering
\begin{tabular}{|ccccc|}  \hline \hline
Set & $\hat{s}^2_Z$ & $\alpha_s (M_Z)$ & $m_t$ (GeV) & $\Delta
\chi^2_H$ \\ \hline
All indirect & $0.2317 (3)(2)$ & $0.124 (5)(2)$ & $173{\pm 11
}^{+17}_{-19}$ & 3.3 \\
Indirect $+$ CDF (174 $\pm$ 16)    & $0.2317 (3)(3)$ & $0.124
(5)(2)$ & $174\pm 9 \pm 12     $ & 3.0 \\
LEP $+$ low energy  & $0.2321 (4)(2)$ & $0.126 (5)(2)$ &
$165^{+11 \; +17}_{-12 \; -19}$ & 1.6 \\
All indirect $(S=3.1)$ & $0.2319 (4)(2)$ & $0.125 (5)(2)$ &
$169^{+11 \; +17}_{-12 \; -19}$ & 2.5 \\
$Z$-pole               & $0.2316 (4)(2)$ & $0.124 (5)(2)$ &
$178^{+11 \; +17}_{-12 \; -19}$ & 3.1 \\
LEP                    & $0.2320 (4)(2)$ & $0.126 (5)(2)$ &
$168^{+12 \; +17}_{-13 \; -19}$ & 1.5 \\
SLD $+ \; M_Z$      & $0.2291 (10)(0)$ & ---            &
$244^{+23 \; +19}_{-25 \; -22}$ & \   \\ \hline
\end{tabular}
\caption{Results for the electroweak parameters in the standard model from
various sets of data.  The central values assume $M_H = 300$~GeV, while the
second errors are for $M_H \ra 1000 (+)$ and $60(-)$.  The last column is
the increase in the overall $\chi^2$ of the fit as $M_H$ increases from 60
to 1000. $m_t$ would increase by some 3 GeV in the fits to the indirect
data if one included the estimates of the $\alpha \, \alpha^2_s \, m_t^2$
threshold corrections.}
\label{tab4}
\end{table}
The first row of the table includes the global fit to all indirect data.
The predicted value of $m_t$ is in remarkable agreement with the value $174
\pm 16$~GeV suggested by the CDF candidate events \cite{a2b}.  The second
row includes the direct (CDF) value for $m_t$ as a separate constraint.
The other fits show the sensitivity to the various data sets.  The third
row includes the LEP results and the low energy data but not SLD.
Comparing with the first row one sees that the predicted $m_t$ is pulled up
significantly (by $\sim 10$~GeV) by the SLD result.  The next row combines
the LEP and SLD measurements of $A_e$ using the scale factor of $3.1$.  The
last rows are the result of the $Z$-pole, LEP, and SLD observables by
themselves.

Using the results of the 1993 LEP energy scan we can now extract the strong
coupling constant $\alpha_s$ at the $Z$-pole with a small experimental
error,
\beq \alpha_s (M_Z) = 0.124 \pm 0.005 \pm 0.002 \;\;\;\; {\rm
(lineshape)},\label{eq30a} \eeq
where the second uncertainty is from $M_H$. $\alpha_s$ is almost
uncorrelated with the other parameters.  It is determined mainly from the
ratio $R \equiv \Gamma (\rm had)/ \Gamma (\ell \bar{\ell})$, which is
insensitive to $m_t$ (except in the $b \bar{b}$ vertex), and also from
$\Gamma_Z$.  This determination is very clean theoretically, at least
within the standard model.  It is the $Z$-pole version of the long held
view that the ratio of hadronic to leptonic rates in $e^+e^-$ would be a
``gold plated'' extraction of $\alpha_s$ and test of QCD.  Using a recent
estimate \cite{a21} of the $(\alpha_s/\pi)^4$ corrections to $C_F$, {\it
i.e.} $- 90 (\alpha_s/\pi)^4$, one can estimate that higher-order terms
lead to an additional uncertainty $\sim \pm 0.001$ in the $\alpha_s(M_Z)$
value in (\ref{eq30a}).  It should be cautioned, however, that the
lineshape value is rather sensitive to the presence of some types of new
physics.

The lineshape value of $\alpha_s$ is an excellent agreement with the
independent value $\alpha_s (M_Z) = 0.123 \pm 0.005$ extracted from jet
event shapes at LEP using resummed QCD \cite{a22}.  It is also in
excellent agreement with the prediction
\beq \alpha_s (M_Z) \sim 0.127 \pm 0.008 ,\;\;\;\;\; {\rm
SUSY-GUT} \eeq
of supersymmetric grand unification.  As can be seen in Table~\ref{tab5},
however, it is somewhat larger than some of the low energy determinations
of $\alpha_s$ (which are then extrapolated theoretically to the $Z$-pole),
in particular those from deep inelastic scattering and the lattice
calculation of the charmonium spectrum\footnote{The lattice value
$0.110 \pm 0.006$ \cite{a22a} has increased somewhat from the published value
of $0.105 \pm 0.004$ \cite{a22b}, reducing the discrepancy.}.
This slight discrepancy has led
some authors to suggest that there might be a light gluino which would
modify the running of $\alpha_s$.  I think, however, that it is premature
to draw such a strong conclusion.  It should be noted that there is an
independent low energy LEP determination from the ratio $R_\tau$ of
hadronic to leptonic $\tau$ decays, which gives a larger value.

\begin{table}                       \centering
\begin{tabular}{|l|c|} \hline \hline
Source & $\alpha_s (M_Z)$ \\ \hline
$R_\tau$ & $0.122 \pm 0.005$ \\
Deep inelastic & $0.112 \pm 0.005$ \\
$\Upsilon$, $J/\Psi$ & $ 0.113 \pm 0.006$ \\
Charmonium spectrum (lattice) & $0.110 \pm 0.006$ \\
LEP, lineshape & $0.124 \pm 0.005 \pm 0.002$ \\
LEP, event topologies & $0.123 \pm 0.005$ \\ \hline
\end{tabular}
\caption{Values of $\alpha_s$ at the $Z$-pole extracted from
various methods.}
\label{tab5}
\end{table}

\subsection{The Higgs Mass}

The new data also constrain  the Higgs boson mass.  This enters
$\hat{\rho}$ logarithmically and is strongly correlated with the
quadratic $m_t$   dependence in everything but  the $Z \ra b \bar{b}$
vertex correction.  The $\chi^2$ distribution as a function of the Higgs
mass is shown in Figure~\ref{fig2}; the minimum occurs at the lower limit,
60 GeV, allowed by direct searches at LEP, or at $\sim 120$~GeV when the CDF
$m_t$ value is included.  These low values are consistent with the minimal
supersymmetric extension of the standard model, which generally predicts a
relatively light standard model-like Higgs scalar.  However, the constraint
is very weak statistically.  From the $\chi^2$ distribution one obtains the
weak upper limits
\beq {\rm indirect: \;\;\;} M_H < 780 (1160) {\rm GeV} \eeq
at 90 (95)\% CL from the indirect precision data, and
\beq {\rm indirect+CDF\; : \;\;\;} M_H < 740 (1040) {\rm GeV} \eeq
including the CDF direct constraint from $m_t$.  (These results
include the direct limit $M_H > 60$~GeV.)  Clearly, no definitive
conclusion can be drawn.
\begin{figure}
\postscript{/home/pgl/fort/nc/graph/chis/xxhiggs.ps}{0.6}
\caption{ $\chi^2$ distributions of the overall fits as a
function of $M_H$.}
\label{fig2}
\end{figure}
An additional strong caveat is in order: the preference for small $M_H$ is
driven almost entirely by $\Gamma(b\bar{b})$, which is significantly above
the standard model prediction even for $M_H = 60$~GeV.  If that is due to a
large statistical fluctuation or to some new physics then the constraint on
$M_H$ would essentially disappear.  Finally, the statistical significance
of the result would decrease even more if the SLC result were omitted.

The weak $M_H$ dependence does not imply that the data is insensitive to
the spontaneous symmetry breaking mechanisms.  Alternative schemes
generally yield large effects on the precision observables,
as will be described below.

\subsection{Have Electroweak Corrections Been Seen?}

The data can also be interpreted in terms of whether one has actually
observed the electroweak (as opposed to the simple running $\alpha$)
corrections.  Novikov {\it et al}. \cite{a15} have noted that there is a
large cancellation between the fermionic and bosonic contributions to the
$W$ and $Z$ self-energies, and that until the most recent data the data
could actually be fit by a properly interpreted Born theory.  However, the
data is now sufficiently good that even given the cancellations these
electroweak loops are needed at the $2\sigma$ level.  Gambino and Sirlin
\cite{a23} and Schildknecht \cite{a24} have interpreted the data in
somewhat different way.  They have argued that the fermionic loops,  both
in the running of $\alpha$ and the $t,b$ loops, are unambiguous
theoretically, and certainly should be there if the theory is to make any
sense. However, the bosonic loops, which involve triple-gauge vertices,
gauge-Higgs vertices, etc., have never been independently tested in other
processes.  They have shown that the data are inconsistent if one simply
ignores bosonic loops (which are a gauge-invariant subset of diagrams),
thus providing convincing though indirect evidence
for their existence.

\section{New Physics}
\subsection{Supersymmetry and Precision Experiments}

Let us now consider how the predictions for the precision observables are
modified in the presence of supersymmetry.  There are basically three
implications for the precision results.  The first, and most important, is
in the Higgs sector.  In the standard model the Higgs mass is arbitrary.  It
is controlled by an arbitrary quartic Higgs coupling, so that $M_H$ could
be as small as 60 GeV (the experimental limit) or as heavy as a TeV.  The
upper bound is not rigorous: larger values of $M_H$ would correspond to
such large quartic couplings that perturbation theory would break down.
This cannot be excluded, but would lead to a theory that is qualitatively
different from the (perturbative) standard model.  In particular, there are
fairly convincing triviality arguments, related to the running of the
quartic coupling, which exclude a Higgs which acts like a distinct elementary
particle for $M_H$ above $O(600$~GeV) \cite{a25}.

However, in supersymmetric extensions of the standard model the quartic
coupling is no longer a free parameter.  It is given by the squares of
gauge couplings, with the result that all supersymmetric models have at
least one Higgs scalar that is relatively light, typically with a mass
similar to the $Z$ mass.  In the minimal supersymmetric standard model
(MSSM) one has $M_H < 150$~GeV\footnote{At tree-level,
$M_H < M_Z$.}, which generally acts just like the standard
model Higgs\footnote{This is true if the second Higgs doublet is much
heavier than $M_Z$.} except that it is necessarily light.

In the standard model there is a large $m_t - M_H$ correlation, and one
has the prediction
\beq m_t \sim 173 \pm 11 + 13 \ln \left( \frac{M_H}{300 \rm
GeV} \right).\eeq
We have seen that for $60< M_H < 1000$~GeV this corresponds to
\beq m_t = 173 \pm11 ^{+17}_{-19}\  ({\rm SM}). \eeq
However, in MSSM one has the smaller range $60 < M_H < 150$~GeV, leading to
the lower prediction
\beq m_t = 159^{+11}_{-12} \pm 5 \ (\rm MSSM).\eeq
This is on the low side of the CDF range, $(174 \pm 16$~GeV), but not
 excluded.

There can be additional effects on the radiative corrections due to
sparticles and the second Higgs doublet that must be present in the MSSM.
However, for most of the allowed parameter space one has $M_{\rm new}
\gg M_Z$, and
the effects are negligible by the decoupling theorem.  For example, a large
$\tilde{t} - \tilde{b}$ splitting would contribute to the $\rho_0$
($SU_2$-breaking) parameter (to be discussed below), leading to a smaller
prediction for $m_t$, but these effects are negligible for $m_{\tilde{q}}
\gg M_Z$.  Similarly, there would be new contributions to the $Z\ra
b\bar{b}$ vertex for $m_{\chi^\pm}$, $m_{\tilde{t}}$, or $M_H^\pm \sim M_Z$.

There are only small windows of allowed parameter space for which the new
particles contribute significantly to the  radiative corrections.  Except
for these, the only implications of supersymmetry from the precision
observables are: (a) there is a light standard model-like Higgs, which in
turn favors a smaller value of $m_t$.  Of course, if a light Higgs were
observed it would be consistent with supersymmetry but would not by itself
establish it.  That would require the direct discovery of the
superpartners, probably at the LHC.  (b) Another important implication of
supersymmetry, at least in the minimal model, is the {\it absence} of other
deviations from the standard model predictions.  (c) In supersymmetric
grand unification one expects the gauge coupling constants to unify when
extrapolated from their low energy values \cite{a27}.  This is consistent
with the data in the MSSM but not in the ordinary standard model (unless
other new particles are added).  This is not actually a modification of the
precision experiments, but a prediction for the observed gauge couplings.
Of course, one could have supersymmetry without grand unification.

\subsection{Extended Technicolor/Compositeness}

In contrast, the other major class  of extensions, which includes
compositeness and dynamical symmetry breaking, leads to many implications
at low energies.  The most important are large flavor changing neutral
currents (FCNC).  Even if these are somehow evaded one generally expects
anomalous contributions to the $Z \ra b\bar{b}$ vertex, typically
$\Gamma(b\bar{b}) < \Gamma^{SM}(b\bar{b})$ in the simplest extended
technicolor (ETC) models \cite{a28}.  Similarly, one expects $\rho_0 \neq
1$, and $S_{\rm new} \neq 0, T_{\rm new} \neq 0$, where $\rho_0$, $S_{\rm
new}$, and $T_{\rm new}$ parameterize certain types of new physics, as will
be described below.  Finally, in theories with composite fermions one
generally expects new 4-fermi operators generated by constituent
interchange, leading to effective interactions of the form
\beq L = \pm \frac{4 \pi}{\Lambda^2} \bar{f}_1 \Gamma f_2 \bar{f}_3
\hat{\Gamma}f_4.\eeq
Generally, the $Z$-pole observables are not sensitive to such operators,
since they only measure the properties of the $Z$ and its
couplings\footnote{At the $Z$-pole the effects of new operators are out
of phase with the $Z$ amplitude and do not interfere. Interference
effects can survive away from the pole, but there
the $Z$ amplitude is smaller.}.
However, low energy experiments are sensitive.  In particular, FCNC
constraints typically set limits of order $\Lambda \geq O(100\; {\rm TeV})$
on the scale of the operators unless the flavor-changing effects are
fine-tuned away.  Even then there are significant limits from other flavor
conserving observables.  For example, atomic parity violation \cite{a29} is
sensitive to operators such as \cite{a30}
\beq L = \pm \frac{4\pi}{\Lambda^2} \bar{e}_L \gamma_\mu e_L
\bar{q}_L \gamma^\mu q_L .\eeq
The existing data already sets limits $\Lambda > O(10$~TeV).  Future
experiments should be sensitive to $\sim$ 40~TeV.

\subsection{The $Zb\bar{b}$ Vertex}

The $Zb\bar{b}$ vertex is especially interesting, both in the standard model
and in the presence of new physics.  In the standard model there are
special vertex contributions which depend quadratically on the top quark
mass, which are shown approximately in (\ref{eq26a}). $\Gamma (b\bar{b})$
actually decreases with $m_t$, as opposed to other widths which all
increase due to the $\hat{\rho}$ parameter.  The $m_t$ and $M_H$
dependences in $\hat{\rho}$ are strongly correlated, but the special vertex
corrections to $\Gamma(b\bar{b})$ are independent of $M_H$, allowing a
separation of $m_t$ and $M_H$ effects.


The vertex is also sensitive to a number of types of new physics.
One can parameterize such effects by \cite{a30a}
\beq \Gamma {(b\bar{b})} \ra \Gamma^{SM} {(b\bar{b})} \left( 1 +
\delta^{\rm new}_{bb} \right) \sim \Gamma^0 {(b\bar{b})} \left( 1
+ \delta^{\rm SM}_{bb} + \delta^{\rm new}_{bb} \right).
\eeq
If the new physics gives similar contributions to vector and axial vector
vertices then the effects on $A_{\rm FB}^{b}$ are negligible.  In
supersymmetry one can have both positive and negative contributions
\cite{a31}.  In particular, light $\tilde{t} - \chi^{\pm}$ can give
$\delta^{\rm SUSY}_{bb} > 0$, as is suggested by the data, while light
charged Higgs particles can yield $\delta^{\rm Higgs}_{bb} < 0$.  In
practice, both effects are too small to be important in most allowed
regions of parameter space.  In extended technicolor (ETC) models there are
typically new vertex contributions generated by the same ETC interactions
which are needed to generate the large top quark mass.  It has
been argued that these are typically large and negative \cite{a28},
\beq \delta^{\rm ETC}_{bb} \sim - 0.056 \xi^2 \left(
\frac{m_t}{150{\rm GeV}} \right),\eeq
where $ \xi$ is a model dependent parameter of order unity.
They may be smaller in models with walking technicolor, but nevertheless
are expected to be negative and significant \cite{a33}.  This is in
contrast to the data, which suggests a positive contribution if any,
implying a serious problem for many ETC models.  One possible way out are
models in which the ETC and electroweak groups do not commute, for which
either sign is possible \cite{a34}.

Another possibility is mixing  between the $b$ and exotic heavy fermions
with non-canonical weak interaction quantum numbers.  Many extensions of
the standard model predict, for example, the existence of a heavy
$D_L$, $D_R$, which are both $SU_2$ singlet quarks with charge $-1/3$.
These can mix with the $d$,
$s$, or $b$ quarks, but one typically expects such mixing to be largest for
the third generation.  However, this mechanism gives a negative
contribution
\beq \delta_{bb}^{D_L} \sim -2.3 s^2_L \eeq
to $\delta^{\rm new}_{bb}$, where $s_L$ is the sine of the $b_l - D_L$
mixing angle.

One can extract $\delta^{\rm new}_{bb}$ from the data, in a global fit to
the standard model parameters as well as $\delta^{\rm new}_{bb}$.  This
yields
\beq \delta^{\rm new}_{bb} = 0.031 \pm 0.014,  \eeq
which is $\sim 2.2\sigma$ above zero.  This value is hardly changed when
one allows additional new physics, such as described by the $S$, $T$, and
$U$ parameters.  Note that $\delta^{\rm new}_{bb}$ is correlated with
$\alpha_s(M_Z)$: one obtains $\alpha_s(M_Z) = 0.103 \pm 0.011$,
considerably smaller than the standard model value $0.124 (5)(2)$. Allowing
$\delta_{bb}^{\rm new} \neq 0$ has negligible effect on $\hat{s}^2_Z$ or
$m_t$.

\subsection{$\rho_0$: Nonstandard Higgs  or Non-degenerate Heavy
Multiplets}

One parameterization of certain new types of physics is the parameter
$\rho_0$, which is introduced to describe new sources of $SU_2$ breaking
other than the ordinary Higgs doublets or the top/bottom splitting.
New physics can affect $\rho_0$ at either the tree or loop-level
\beq \rho_0 = \rho_0^{\rm tree} + \rho_0^{\rm loop}.\eeq
The tree-level contribution is given by Higgs representations
larger than doublets, namely,
\beq \rho_0^{\rm tree} = \frac{\sum_i \left( t^2_i -  t_{3i}^2
+ t_i \right) |\langle \phi_i \rangle|^2}{ \sum_i 2 t_{3i}^2
|\langle \phi_i \rangle|^2},  \label{eqerica}\eeq
where $t_i$ ($t_{3i}$) is the weak isospin (third component) of the
neutral Higgs field $\phi_i$.
If one has only Higgs singlets and doublets ($t_i = 0,\frac{1}{2}$),
then $\rho_0^{\rm tree} = 1$.
However, in the presence of larger representations with non-zero
vacuum expectation values
\beq \rho_0^{\rm tree} \simeq 1 + 2 \sum_i \left( t^2_i -
3 t_{3i}^2 + t_i \right) \frac{ |\langle \phi_i \rangle
|^2}{|\langle \phi_{\frac{1}{2}} \rangle |^2 }. \label{eq22a3}
\eeq

One can also have loop-induced contributions similar to that of the
top/bottom, due to non-degenerate multiplets of fermions or bosons.  For new
doublets
\beq \rho_0^{\rm loop} = \frac{3G_f}{8 \sqrt{2} \pi^2} \sum_i
\frac{C_i}{3} F (m_{1i},m_{2i}),   \eeq
where $C_i = 3(1)$ for color triplets (singlets) and
\beq F(m_1, m_2) = m_1^2 + m^2_2 - \frac{4m_1^2 \; m^2_2}{m_1^2 -
m^2_2} \ln \frac{m_2}{m_2} \geq (m_1- m_2)^2 . \eeq
Loop contributions to $\rho_0$ are generally positive,\footnote{One can
have $\rho^{\rm loop} < 0$ for Majorana fermions \protect\cite{a34a} or
boson multiplets with vacuum expectation values \protect\cite{a34b}.} and
if present would lead to lower values for the predicted $m_t$. $\rho_0^{\rm
tree}$ can be either positive or negative depending on the quantum numbers
of the Higgs field.  The $\rho_0$ parameter is extremely important because
one expects $\rho_0 \sim 1$ in most superstring theories, which generally
do not have higher-dimensional Higgs representations, while typically
$\rho_0 \neq 1$ from many sources in models involving compositeness.

In the presence of $\rho_0$ the standard model formulas for the observables
are modified.  One has
\beq M_Z \ra \frac{1}{\sqrt{\rho_0}} M_Z^{SM}, \Gamma_Z \ra
\rho_0
\Gamma_Z^{SM}, {\cal{L}}_{NC} \ra \rho_0 {\cal{L}}^{SM}_{NC}.\eeq
It has long been known that $\rho_0$ is close to 1. However, until recently it
has been difficult to separate $\rho_0$ from $m_t$, because in most
observables one has only the combination $\rho_0 \hat{\rho}$.  The one
exception has been the $Z \ra b\bar{b}$ vertex.  However, assuming that CDF
has really observed the top quark directly one can use the known $m_t$ to
calculate $\hat{\rho}$ and therefore separate $\rho_0$.  In practice one
fits to $m_t$, $\rho_0$ and the other parameters, using the CDF value $m_t
= 174 \pm 16$~GeV as an additional constraint.  One can determine
$\hat{s}^2_Z$, $\rho_0$, $m_t$, and $\alpha_s$ simultaneously, yielding
\beq \begin{array}{ccc} \hat{s}^2_Z = 0.2316(3)(2) & \;\;\; & \rho_0 = 1.0009
\pm 0.0018 \pm 0.0017 \\ \alpha_s = 0.123(6)(1) & & m_t = 167 \pm 15 \pm 1 \
{\rm GeV}, \end{array} \label{eqrho} \eeq
where the second uncertanty is from $M_H$.  Even in the presence of the
classes of new physics parameterized by $\rho_0$ one still has robust
predictions for the weak angle and a good determination of $\alpha_s$.
Most remarkably, given the CDF constraint, $\rho_0$ is constrained to be
very close to unity, causing serious problems
for compositeness models.  The allowed region in $\rho_0$ vs $\hat{s}^2_Z$ are
shown in Figure \ref{figerica}.  This places limits $|\langle \phi_i
\rangle|/ |\langle \phi_{1/2} \rangle| < {\rm few} \%$ on non-doublet
vacuum expectation values, and places constraints $\frac{C}{3} F(m_1, m_2)
\leq (100\; {\rm GeV})^2$ on the splittings of additional fermion or boson
multiplets.

\begin{figure}
\postscript{/home/pgl/fort/nc/graph/xrho/xxrhof5.ps}{0.6}
\caption{Allowed regions in $\rho_0$ vs $\hat{s}^2_Z$ for $M_H = 60$, 300,
and 1000 GeV.}
\label{figerica}
\end{figure}

\subsection{Heavy Physics by Gauge Self Energies}
\label{stusec}

A larger class of extensions of the standard model can be parameterized by
the $S$, $T$ and $U$ parameters \cite{a35}, which describe that subset of
new physics which affect only the gauge boson self-energies but do not
directly affect new vertices, etc.  One introduces three parameters
\begin{eqnarray} S &=& S_{\rm new} + S_{m_t} + S_{M_H} \nonumber \\
                 T &=& T_{\rm new} + T_{m_t} + T_{M_H}  \\
   U &=& U_{\rm new} + U_{m_t}. \nonumber \end{eqnarray}
$S$ describes the breaking of the $SU_{2A}$ axial generators and is
generated, for example, by degenerate heavy chiral families of fermions.
$T$ and $U$ describe the breaking of $SU_{2V}$ vector generators:
 $T$ is equivalent to the
$\rho_0$ parameter and is induced by mass splitting in multiplets of
fermions or bosons. $U$ is zero in most extensions of the standard model. $S$,
$T$ and $U$ were introduced to describe the contributions of new physics.
However, they can also parametrize the effects of very heavy $m_t$ and
$M_H$ (compared to $M_Z$).  Until recently it was difficult to separate the
$m_t$ and new physics contributions.  Now, however, with the CDF value of
$m_t$ it is possible to directly extract the new physics contributions.

A new multiple of degenerate chiral fermions will contribute to $S_{\rm
new}$ by
\beq S_{\rm new} |_{\rm degenerate} = C_i |t_{3L} (i) - t_{3R} (i)
|^2/3\pi \geq 0, \eeq
where $C_i$ is the number of colors and $t_{3LR}$ are the $t_3$ quantum
numbers.  A fourth family of degenerate fermions would yield
$\frac{2}{3\pi} \sim 0.21$, while QCD-like technicolor models, which
typically have many particles, can give larger contributions.  For example,
$S_{\rm new} \sim 0.45$ from an isodoublet of fermions
with four technicolors, and an
entire technigeneration would yield $1.62$ \cite{a36}.  Non-QCD-like theories
such as those involving walking could yield smaller or even
negative contributions \cite{a37}.  Nondegenerate scalars or fermions can
contribute to $S_{\rm new}$ with either sign \cite{a38}.
(Note that $S$, $T$, and
$U$ are induced by loop corrections and have a factor of $\alpha$ extracted,
so they are expected to be $O(1)$ if there is new physics.)

The $T$ parameter is analogous to $\rho_0^{\rm loop}$.  For a
non-degenerate family
\beq T_{\rm new} \sim \frac{\rho_0^{\rm loop}}{\alpha} \sim 0.42
\frac{\Delta m^2}{(100 \;GeV)^2}, \eeq
where
\beq \Delta m^2 = \sum_i \frac{C_i}{3} F \left( m_{1i}, m_{2i} \right) \geq
\sum_i \frac{C_i}{3 } \left( m_{1i} - m_{2i} \right)^2.\eeq
Usually $T_{\rm new} > 0$, although there may be exceptions for theories
with Majorana fermions or additional Higgs doublets.  In practice,
higher-dimensional Higgs multiplets could mimic $T_{\rm new}$ with either
sign (see equation (\ref{eqerica})), and cannot be separated from loop
effects unless they are seen directly or have other effects.  Usually
$U_{\rm new}$ is small.

There is enough data to simultaneously determine the new physics
contributions to $S$, $T$, and $U$, the standard model parameters, and also
$\delta^{\rm new}_{bb} =
\frac{\Gamma (b\bar{b})}{\Gamma^{\rm SM} (b\bar{b})} -1
$. For example, $S_{\rm new}$, $T_{\rm new}$, $U_{\rm new}$, $\delta^{\rm
new}_{bb}$, $\hat{s}^2_{Z}$, $\alpha_s(M_Z)$ and $m_t$ are constrained by
$M_Z$, $\Gamma$, $M_W$, $R_b$, asymmetries, $R$, and $m_t$ (CDF),
respectively.  One obtains
\begin{eqnarray} S_{\rm new} = -0.15 \pm 0.25^{-0.08}_{+0.17} &
\; \; \;\;\; & \hat{s}^2_Z = 0.2314 (4) \nonumber \\
T_{\rm new} = -0.08 \pm 0.32^{+0.18}_{-0.11} &
\; \; \;\;\; & \alpha_s(M_Z) = 0.103 (11) \nonumber \\
U_{\rm new} = -0.56 \pm 0.61 & \; & m_t = 175 \pm 16 \; {\rm GeV}
 \\
\delta^{\rm new}_{bb} = 0.031 \pm 0.014, & \; & \; \nonumber
\end{eqnarray}
where the second error is from $M_H$.
The $T_{\rm new}$ value corresponds to
$\rho_0 = 0.9994 \pm 0.0023 ^{+ 0.0013}_{- 0.0008}$, which differs
from the value in (\ref{eqrho}) because of the presence of $S_{\rm new}$,
$U_{\rm new}$, and $\delta^{\rm new}_{bb}$.
The data is consistent with the
standard model: $S_{\rm new}$ and $T_{\rm new}$ are close to zero with
small errors, and the tendency to find $S < 0$ that existed
in earlier data is no longer present.
The constraints on $S_{\rm new}$ are a problem for those
classes of new physics such as technicolor which tend to give $S_{\rm new}
$ large and positive, and $S_{\rm new}$ allows, at most, one additional
family of ordinary fermions at 90\% CL.  (Of course the invisible $Z$ width
precludes any new families unless the additional neutrinos are heavier than
$M_Z/2$.)  The allowed regions in $S'_{\rm new}$ vs $T'_{\rm new}$ (which
include the very small Higgs contributions) are shown in Figure~\ref{fig5}.
The seven parameter fit still favors a non-zero $Z\ra b\bar{b}$ vertex
correction $\delta^{\rm new}_{bb}$.

The value of $\hat{s}^2_Z$ is slightly lower than the standard model value
$(0.2317 (3)(2))$.  However, the extracted $\alpha_s(M_Z)$ is considerably
lower than the standard model value $(0.124 (5)(2))$.  This is entirely due
to the presence of $\delta^{\rm new}_{bb}$. By allowing $\delta^{\rm
new}_{bb} > 0$ one can describe $R = \Gamma ({\rm had})/\Gamma(\ell
\bar{\ell})$ with a smaller QCD correction to $\Gamma(\rm had)$.  Thus,
$\alpha_s(M_Z)$ from the lineshape, though very clean in the standard
model, is more sensitive to certain types of new physics than most other
determinations.

\begin{figure}[tbh]
\postscript{/home/pgl/fort/nc/graph/st/st5b.ps}{0.7}
\caption{Constraints on $S'_{\rm new}$ and $T'_{\rm new}$ from various
observables and from the global fit to all data, where $S'_{\rm new} =
S_{\rm new} + S_{M_H}$, and similarly for $T'_{\rm new}$.  The circle,
square, and diamond represent the standard model expectations for $M_H
=$~60, 300, and 1000, respectively.}
\label{fig5}
\end{figure}

\section{Conclusions}

\begin{itemize}

\item The precision data have confirmed the standard electroweak model.
However, there are possible hints of discrepancies at the 2 -- 3 $\sigma$
level in $\Gamma (b\bar{b})/\Gamma(\rm had)$ and $A^0_{LR}$.

\item The data not only probes the tree-level structure, but the
electroweak loops have been observed at the $2\sigma$ level.  These consist
of much larger fermionic pieces involving the top quark and QED, which only
partially cancel the bosonic loops.  The bosonic loops, which probe
non-abelian vertices and gauge-Higgs vertices, are definitely needed to
describe the data.

\item The global fit to the data within the standard model yields
\beq \begin{array}{ccc} \overline{MS}: \hat{s}^2_Z = 0.2317 (3)(2) &
\;\;\;\; & m_t = 173 \pm 11\,^{+17}_{-19}\\ {\rm on-shell:} \; s^2_W
\equiv 1 - \frac{M_W^2}{M_Z^2}  = 0.2242 (12) & & \alpha_s (M_Z)
= 0.124 (5)(2), \end{array} \eeq
where the second uncertainty is from $M_H$.  The prediction for $m_t$ is in
remarkable agreement with the value $m_t = 1 74 \pm 16$ suggested by the
CDF events.  The data has also allowed, for the first time, a clean and
precise extraction of $\alpha_s$ from the lineshape.  This is in excellent
agreement with the value $\alpha_s (M_Z) = 0.123 \pm 0.005$ from event
shapes.  Both are larger than many of the low energy determinations when
extrapolated to the $Z$-pole.  The lineshape determination, however, is
sensitive to the presence of certain types of new physics.

\item The agreement between the indirect prediction for $m_t$
with the tentative direct CDF observation
and of $\alpha_s$ with the various other determinations
is an impressive success for the
entire program of precision observables.

\item Combining the direct CDF value of $m_t$ with the indirect constraints
does not make a large difference within the context of the standard model.
However, when one goes  beyond the standard model, the direct $m_t$
allows a clean extraction of the new physics contributions to $\rho_0$,
which is now shown to be very close to unity, $\rho_0 = 1.0009 (18)(17)$.
This strongly limits Higgs triplet vacuum expectation values and
non-degenerate heavy multiplets.  Similarly, it allows an extraction of the
new physics contributions to $S_{\rm new}$, $T_{\rm new}$, $U_{\rm new}$,
which are consistent with zero.  Finally, one can determine the new physics
contributions to the $b\bar{b}$ vertex: $\delta^{\rm new}_{bb}$ is
approximately $2.2\sigma$ away from zero, reflecting the large value of the
$b\bar{b}$ width.

\item The data exhibit a slight preference for a light Higgs, but this is
not very compelling statistically.  One finds only $M_H \leq 780 (1160)$~GeV
at 90(95\%) CL.  Furthermore, the preference depends crucially on the large
observed value of $\Gamma(b\bar{b})$, and to a lesser extent on the
large SLD value for $A^0_{LR}$.  Omitting these
values the  $M_H$ dependence of the observables is weak.

\item The major prediction of supersymmetry is that one does not expect
large deviations in the precision observables.  The new particles tend to
be heavy and decouple.  One implication that is relevant, however, is that
supersymmetric theories have a light standard model-like Higgs.  They
therefore favor the lighter Higgs mass and the lower end of the predicted
$m_t$ range.  Also, the observed gauge couplings are consistent with the
coupling constant unification expected in supersymmetric grand unification,
but not with the simplest version of non-supersymmetric unification.

\item In compositeness and dynamical symmetry breaking theories one
typically expects not only large flavor changing neutral currents but
significant deviations of $\rho_0$ from unity and of $S_{\rm new}$ and
$T_{\rm new}$ from zero.  One further expects that $\delta_{bb}^{\rm new} <
0$, at least in the simplest models.  Therefore, the precision experiments
are a major difficulty for this class of models.

\end{itemize}

\section*{Acknowledgement}

It is a pleasure to thank Jens Erler for collaboration on these analyses.
I would also like to thank the conference organizers for travel and
support.

\end{document}